\documentclass[twocolumn,showpacs,preprintnumbers,amsmath,amssymb]{revtex4}

\usepackage{graphicx}
\usepackage{dcolumn}
\usepackage{bm}
\usepackage{amsmath}
\usepackage{times}
\usepackage[usenames]{color}

\begin{document}

\title{Photo-induced spin-state change in itinerant correlated-electron system}
\author{Y.~Kanamori$^1$, H.~Matsueda$^{2}$, and S. Ishihara$^{1, 3}$} 
\affiliation{$^1$Department of Physics, Tohoku University, Sendai 980-8578, Japan}
\affiliation{$^2$Sendai National College of Technology, Sendai, 989-3128, Japan}
\affiliation{$^3$Core Research for Evolutional Science and Technology (CREST), Sendai 980-8578, Japan}

\date{\today}
\begin{abstract}
Photo-induced spin-state change in itinerant correlated-electron system is studied. 
The model Hamiltonians before and after photon-pumping are derived from the two-orbital Hubbard model with crystalline field splitting. 
A photon introduced in the low-spin band insulator induces a bound state of the high-spin state and a photo-excited hole. 
This bound state brings a characteristic peak in the optical absorption spectra in the photo-excited state. 
The present results well explain the recent experimental results of the ultrafast optical spectroscopy in perovskite cobaltites. 
\end{abstract}

\maketitle

%
\narrowtext
%

A number of electronic phases and the phase transition between them are one of the central issues in correlated electron systems. 
In particular, materials with multi-degrees of freedom, such as charge, spin, orbital and so on, exhibit various exotic phenomena related to the phase transition~\cite{Maekawa}. 
Recently developed ultrafast optical techniques open up a new frontier for research of the phase transition~\cite{Nasu}. 
Irradiation of a pump laser pulse into a system in a vicinity of the phase boundary triggers an abrupt change in the electronic structures. 
This is the so-called photo-induced phase transition (PIPT). 
In contrast to the conventional phase transitions caused by controlling temperature ($T$), doping carriers and so on, photo-induced phase is transient and highly nonequilibrium. 
Photo-induced phenomena in correlated electron systems offer large possibility of new hidden phases~\cite{Ichikawa,Wall}, which do not realize in the thermal equilibrium state, and prompt several theoretical challenges~\cite{Matsueda,Kanamori,Freericks,Eckstein,Moritz,Koshibae,Maeshima}. 

The spin-state transition is one of the targets in recent PIPT studies. 
This is a transition between the states with different magnitudes of the spin-angular moment in transition-metal ions. 
Different spin states are realized owing to a delicate balance of the intra-ion Hund coupling and the crystalline-field splitting. 
Some examples have been seen in the insulating organometallic complexes, such as the Prussian-blue analog, 
where magnitudes of the localized spins in Co or Fe ions are switched by photon irradiation~\cite{Gutlich, Decurtins}. 
Another type of the photo-induced spin-state transition is suggested in the correlated electron systems, the cobalt oxides with a perovskite structure, $R_{1-x}$$A_{x}$CoO$_3$ and $R$$A$Co$_2$O$_{6-\delta}$ ($R$: a rare-earth ion, $A$: an alkaline-earth ion). 
Possible three spin states in Co$^{3+}$ are the low-spin (LS) in the $(e_g)^0(t_{2g})^6$ configuration, the intermediate-spin (IS) in $(e_g)^1(t_{2g})^5$, and the high-spin (HS) in $(e_g)^2(t_{2g})^4$~\cite{Tokura,Frontera}. 
Temperature induced spin-state transition from the low-$T$ LS insulating state into the high-$T$ HS or IS metallic one is commonly observed in these cobaltites. 
The ultrafast optical measurements in the low-$T$ LS insulators show transient metallic spectra which are completely different from the spectra in the high-$T$ phase~\cite{Okimoto1,Okimoto2,Iwai}. 
The electron conduction and the spin-state are strongly coupled with each other, in highly contrast to the 
spin-crossover organometallic complexes where $d$ electrons usually hop around the neighboring sites. 
Thus, the photo-induced spin-crossover in the itinerant correlated electrons are ought to be examined from the view point of the cross correlated phenomena among multi-degrees of freedom. 

In this Letter, photo-induced spin-state change in itinerant correlated electron systems is studied. 
We derive the model Hamiltonians where numbers of photo-excited electron-hole pair are fixed. 
A bound state of the photo-doped hole and the HS state are created inside of the LS sites. 
This bound state brings about a characteristic peak structure in the optical pump-probe spectra which are completely different from the spectra in thermal-excited states. 
The present theory provides a possible scenario in recently observed photo-induced hidden state in perovskite cobaltites. 

\begin{figure}[t]
\includegraphics[width=\columnwidth,clip]{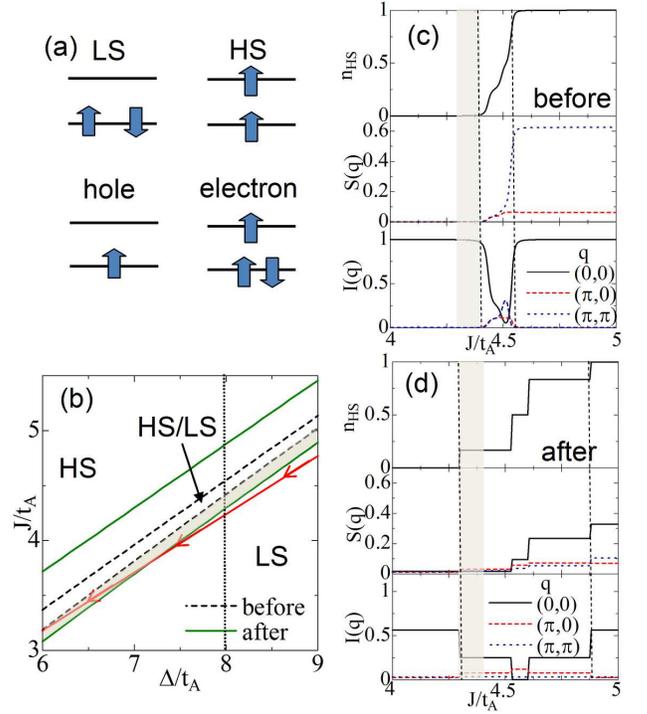}
\caption{(a) schematic configurations of LS, HS, hole and electron states. 
(b) phase diagrams before and after photon pumping. 
A shaded area is for a region where the photo-induced spin-state change from LS to HS occurs. 
A bold line with arrows is for $J=0.53\Delta$. A direction of arrows corresponds to increasing of $t_A$ with fixing $\Delta$ and $J$ (see the text). 
(c) HS density, spin correlation function and spin-state correlation function before photon pumping, 
and (d) those after photon pumping in the case of $\Delta=8t_A$ [a vertical line in (b)]. 
A two dimensional cluster of $N=8$ is adopted.}
\label{fig:fig1}
\end{figure}

We start from the two-orbital Hubbard model with a crystalline field splitting effect as a minimal model to examine the photo-induced spin-state change. 
Two orbitals termed A and B, corresponding to the $e_g$ and $t_{2g}$ orbitals, respectively, are introduced in each site. 
The energy-level difference is denoted as $\Delta=\varepsilon_{ A}-\varepsilon_{ B}>0$. 
The model Hamiltonian in a $N$-site lattice is given by 
\begin{eqnarray}
{\cal H}&=&\Delta \sum_{i \sigma} c_{i A \sigma}^\dagger c_{i A \sigma}
- \sum_{\langle ij \rangle \gamma \sigma} 
t_\gamma
\left ( c_{i \gamma \sigma}^\dagger c_{j \gamma \sigma} +{\rm H.c.} \right ) 
\nonumber \\
&+&U\sum_{i \gamma} n_{i \gamma \uparrow} n_{i \gamma \downarrow}
+U'\sum_{i \sigma \sigma'} n_{i A \sigma} n_{i B \sigma'}
\nonumber \\
&+&J\sum_{i \sigma \sigma'}c_{i A \sigma}^\dagger c_{i B \sigma'}^\dagger c_{i A \sigma'} c_{i B \sigma}
+I\sum_{i \gamma} c_{i \gamma \uparrow}^\dagger c_{i \gamma \downarrow}^\dagger c_{i \bar{\gamma} \downarrow} c_{i {\bar \gamma} \uparrow}, 
\label{eq:ham}
\end{eqnarray}
where $ c_{i \gamma \sigma}$ is the electron annihilation operator at site $i$ with orbital $\gamma(= A, B)$ and spin $\sigma(=\uparrow, \downarrow)$, 
and $n_{i \gamma \sigma}=c^\dagger_{i \gamma \sigma} c_{i \gamma \sigma}$ is the number operator. 
We define ${\bar \gamma}=(A, B)$ for $\gamma=(B, A)$, and introduce the intra-orbital Coulomb interaction $U$, the inter-orbital one $U'$, the exchange interaction $J$, and the pair-hopping $I$. 
The electron transfer $t_\gamma$ is set to be diagonal with respect to the orbitals. 
The total number of electrons is fixed to be $2N$. 
In the case of $t_\gamma=0$, we have the two-lowest electronic configurations, the LS singlet state 
$|\psi_L \rangle=(f_B |B_\uparrow B_\downarrow \rangle - f_A|A_\uparrow A_\downarrow \rangle )$, 
and the HS triplet states 
$\{|\psi_{H0} \rangle, |\psi_{H+1} \rangle, |\psi_{H-1} \rangle \}=\{ (|A_\uparrow B_\downarrow \rangle +|A_\downarrow B_\uparrow \rangle )/\sqrt{2}, |A_\uparrow B_\uparrow \rangle, |A_\downarrow B_\downarrow \rangle \}$, 
as schematically shown in Fig.~\ref{fig:fig1}(a). 
We define the factors  
$f_A=(\sqrt{\Delta^2+I^2}-\Delta)/K$, $f_B=I/K$ and $K=[(\sqrt{\Delta^2+I^2}-\Delta)^2+I^2]^{1/2}$. 
The energies are $E_L=U+\Delta-\sqrt{\Delta^2+I^2}$ for the LS state and $E_H=U'-J+\Delta$ for the HS one. 

We derive the effective Hamiltonians for 
the electronic structures before and after photon-pumping by the perturbational procedure with respect to $t_\gamma$. 
Before photon-pumping, the unperturbed electronic configurations are restricted to be $|\psi_L \rangle $ and $| \psi_{H l} \rangle$ ($l=0, \pm 1$). 
The Hamiltonian is explicitly written by using the projection operators as 
\begin{eqnarray}
&\ &{\cal H}_{0}
=E_L \sum_{i } P^L_i+E_H \sum_{i} P^{H}_i 
+J_{LL}\sum_{\langle ij \rangle}  P^L_{i} P^L_{j} \nonumber \\
&+&J_{HH}\sum_{\langle ij \rangle} 
\left ({\bf S}_i \cdot {\bf S}_j-1 \right ) P^H_i P^H_j 
+J_{LH}\sum_{\langle ij \rangle}  \left (P_{i}^H P_{j}^L+{\rm H.c.} \right ) \nonumber \\
&+&J_{++}\sum_{ \langle ij \rangle} \left \{ \left ({\bf S}_i \cdot {\bf S}_j -1 \right ) P_i^+ P^+_j +{\rm H.c.} \right \}
\nonumber \\
&+&J_{-+}\sum_{\langle ij \rangle} \left \{ P_i^- \left ({\bf S}_i \cdot {\bf S}_j +1 \right ) P^+_j +{\rm H.c.} \right \} ,
\label{eq:h0}
\end{eqnarray}
where $P_i^L = |\psi_{i L} \rangle \langle \psi_{i L}|$, 
$P_i^{H} = \sum_{l=0, \pm 1} |\psi_{i Hl} \rangle \langle \psi_{i Hl}|$ 
and ${\bf S}_i = (1/2) \sum_{l s s'}c^\dagger_{i l s} {\bf \sigma}_{ss'} c_{i l s'}$ with the Pauli matrices ${\bf \sigma }$. 
We also introduce the operators   
$P^+_i=|\psi_{iH0} \rangle \langle \psi_{i L}|$ and 
$P^-_i=(P^+_i)^{\dagger}$ which change the spin state. 
%
The first and second terms are for the on-site energies of the LS and HS states, respectively. 
The remaining terms represent the inter-site interactions.  
Coupling constants are of the order of the $t_\gamma^2/U$, and are explicitly given in Ref.~\cite{exchange}. 
The dominant terms are the nearest neighbor (NN) interaction between LS and HS, i.e. $J_{LH}$ and $J_{-+}$. 
The negative value of $J_{LH}$ (see Ref.~\cite{exchange}) implies an attractive interaction between the LS and HS states~\cite{Khomskii}. 

The effective Hamiltonian for the states after the photon-pumping is derived in a similar way. 
We assume that one photon generates one pair of an electron and a hole. 
These are represented by the configurations of 
$|\psi_{e \sigma} \rangle=| A_\sigma B_\uparrow B_\downarrow \rangle$
and 
$|\psi_{h \sigma} \rangle=| B_{\sigma} \rangle$, 
which are termed an electron site and a hole site, respectively, as shown in Fig.~\ref{fig:fig1}(a).
As the unperturbed states, 
in addition to the LS and HS sites, a pair of electron and hole sites are introduced. 
To avoid redundancy, here we give an outline of the Hamiltonian: 
\begin{eqnarray}
{\cal H}_1={\cal H}^{(e)(h)}+ {\cal H}^{(e/h)(L/H)}+{\cal H}_{0}. 
\end{eqnarray}
The explicit forms will be presented elsewhere. 
The first term represents the NN interactions between photo-excited hole and electron. 
The second term describes the interactions between hole/electron and LS/HS.  
One of the dominant terms in ${\cal H}_1$ is the $t_\gamma$-linear term between the photo-excited hole and HS, e.g.   
$t_A \sum_{\langle ij \rangle} |\psi_{i h \uparrow} , \psi_{j H+1} \rangle  
\langle \psi_{i H+1} ,  \psi_{j h \uparrow} |$. 
When we identify that the A and B electrons are recognized as itinerant electrons and localized spins, respectively, 
this term corresponds to the double-exchange (DE) interaction. 
In the following, we mainly focus on the lowest energy states in ${\cal H}_0$ and ${\cal H}_1$ 
denoted by $|GS_0\rangle$ and $|GS_1 \rangle$, respectively. 
The state $|GS_1 \rangle$ corresponds to the photo-excited state where the energy and the angular moments are fully relaxed within the condition that the photo-excited electron and hole do not disappear. 

The two effective Hamiltonians, ${\cal H}_0$ and ${\cal H}_1$, are analyzed by the exact-diagonalization method with the Lanczos algorithm in finite-size clusters. 
We adopt mainly the two-dimensional 8-site and 10-site clusters with the periodic-boundary condition. 
One dimensional cluster with $N\le 10$ are also used to check the size effect. 
In the finite $T$ calculations, all of the eigen states and the eigen energies are obtained by the Householder algorithm. 
Energy in ${\cal H}_1$ is optimized by the spin indices of the photo-excited hole and electron. 
In the numerical calculations, we set $U=4J$, $U'=3J$, $I=J$ and $t_B=0.1t_A$. 
We have checked that qualitative results do not depend on detailed parameters choice. 

First, we show the electronic structure before photon-pumping. 
The phase diagram in the $\Delta-J$ plane is presented in Fig.~\ref{fig:fig1}(b). 
This is obtained by the HS density  $n_{HS}=(N-2m)^{-1}(\sum_{i}\langle n_{iA} \rangle-m)$, 
the spin correlation function $S({\bf q})=(2N^2)^{-1}\sum_{ij} \langle {\bf S}_i \cdot {\bf S}_j \rangle 
e^{i {\bf q} \cdot ({\bf r}_i-{\bf r}_j)}$, 
and the spin-state correlation function $I({\bf q})=N^{-2}\sum_{ij} \langle Q_i Q_j\rangle 
e^{i {\bf q} \cdot ({\bf r}_i-{\bf r}_j)}$ shown in Fig.~\ref{fig:fig1}(c). 
We define $m=0$ and 1 for ${\cal H}_0$ and ${\cal H}_1$, respectively, 
and $Q_{i}=|\psi_{iL}\rangle \langle \psi_{i L} |-\sum_l |\psi_{i H l}\rangle \langle \psi_{i Hl}|$. 
As expected, the HS and LS phases appear in the regions of large $J$ and large $\Delta$, respectively. 
The antiferromagnetic  spin correlation is developed in the HS phase due to the superexchange interaction $J_{HH}(>0)$ in Eq.~(\ref{eq:h0}).
Between the two phases, LS and HS coexist. 
The staggered spin-state order is shown in the spin-state correlation function $I({\bf q})$. 
This LS-HS order is also confirmed by the calculations in the two-orbital Hubbard model, ${\cal H}$ in Eq.~(\ref{eq:ham}), 
and is attributed to the attractive interaction between LS and HS, i.e. $J_{LH}$ in Eq.~(\ref{eq:h0}). 

The phase diagram after the photon-pumping is also plotted in Fig.~\ref{fig:fig1}(b) 
which is determined from $n_{HS}$, $S({\bf q})$ and $I({\bf q})$ shown in Fig.~\ref{fig:fig1}(d). 
It is clearly shown that the LS-HS coexistence phase becomes wider than that before pumping. 
This phase does not show a remarkable staggered LS-HS correlation and is identified as a metallic state as explained later. 
Equivalently, in the shaded area in Fig.~\ref{fig:fig1}(b), the spin-state change from LS to HS is induced by the photon-pumping.  

\begin{figure}[t]
\includegraphics[width=\columnwidth,clip]{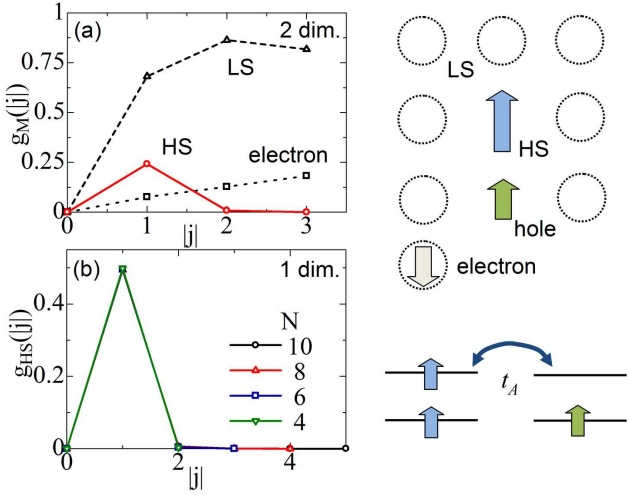}
\caption{(a) spatial distributions around the photo-excited hole in two-dimensional 10-site cluster, 
and (b) those for HS in one-dimensional clusters for several $N$. 
Schematic pictures for the hole-HS bound state are also shown. 
Parameters are $\Delta=8t_A$ and $J=4.325t_A$. }
\label{fig:fig2}
\end{figure}
Properties of the photo-induced HS state are clarified in more detail. 
In Fig.~\ref{fig:fig2}(a), we present the spatial distributions for the local-electronic configurations around the photo-excited hole. 
This is defined by 
$g_{M}(|j|)=\sum_{i } \langle P_{i+|j|}^{M} P_{i}^{ h }\rangle$ 
for $M=(H, L, e)$ where we introduce the projection operators 
$P_{i}^{e }=\sum_{\sigma} | \psi_{i e \sigma} \rangle\langle \psi_{i e \sigma} |$ 
and $P_{i}^{h }=\sum_{\sigma} | \psi_{i h \sigma} \rangle\langle \psi_{i h \sigma} |$.  
It is shown that HS is located at the NN sites of the photo-excited hole. 
This result is almost independent of dimension and size of clusters as shown Fig.~\ref{fig:fig2}(b).
The calculated spin-correlation function between the hole and HS is 0.498 which 
suggests that spins in these sites are polarized ferromagnetically. 
This is a kind of the spin-polarized bound state, like a magnetic polaron, induced by the DE interaction as schematically shown in Fig.~\ref{fig:fig2}.  
We have confirmed that the nature of this bound state is robust against changing the parameters. 

\begin{figure}[t]
\includegraphics[width=\columnwidth,clip]{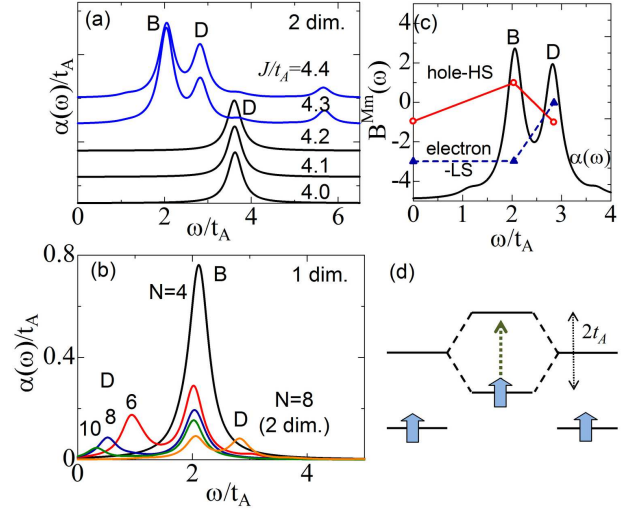}
\caption{(a) the optical absorption spectra calculated in ${\cal H}_1$ for several values of $J/t_A$. 
(b) size dependence of the optical spectra in one-dimensional cluster. 
(c) the bond correlation functions for the final states of the optical absorption spectra. 
The bold and broken lines are for $B^{hH}(\omega)$ and $B^{eL}(\omega)$, respectively. 
The optical spectra are also shown. 
(d) a schematic picture of the optical excitation between the bonding and antibonding orbitals in a hole-HS bound state.
Parameters are $\Delta=8t_A$, $J=4.325t_A$, and $\eta=0.2t_A$. 
A two dimensional $N=8$ cluster is adopted in (a) and (c). 
}
\label{fig:fig3}
\end{figure}

The photo-excited ferromagnetic bound state is able to be detected directly by the optical pump-probe experiments. 
The optical absorption spectra calculated in ${\cal H}_1$ are shown in Fig.~\ref{fig:fig3}(a). 
This is defined by 
$\alpha_{l}(\omega)=-(\pi N)^{-1}{\rm Im}\langle  j_l (\omega-{\cal H}_1+E+ i \eta )^{-1}  j_l  \rangle$
where $j_l=i \sum_{\sigma \gamma i } t_\gamma
(c^\dagger_{i \gamma \sigma} c_{i+l \gamma \sigma}-c^\dagger_{i+l \gamma \sigma} c_{i \gamma \sigma})$
is the current operator along the direction $l$, $E=\langle  {\cal H}_1  \rangle$
and $\eta$ is a small damping constant. 
A symbol $\langle \cdots \rangle$ implies the expectation in terms of $| GS_1 \rangle$. 
%
In the case where the HS state is induced by the photon-pumping $(J/t_A \geq 4.3)$, mainly two peaks, denoted as D and B, appear. 
On the other hand, a single peak (the peak D) is seen in the case where the HS state is not induced $(J/t_A \leq 4.2)$.
Size dependence is checked in one-dimensional clusters [Fig.~\ref{fig:fig3}(b)]. 
The energy of the peak D decreases monotonically and tends to zero with increasing $N$. 
The energy of the peak B is, however, almost independent of dimension and size of clusters, and remains to be $\omega \sim 2 t_A$. 

The origin of the peaks in $\alpha(\omega)$ is more clarified by 
analyses of the final states in $\alpha(\omega)$. 
We introduce the bond-correlation function defined by $B^{mM}(\omega)=\sum_{\langle ij \rangle \sigma \gamma} \langle \psi(\omega) | P_{j}^{m}P_{i}^{M} c_{i\gamma \sigma}^\dagger c_{j \gamma \sigma}P_{j}^{M}P_{i}^{m}+{\rm H.c.} | \psi(\omega) \rangle $ 
where $|\psi(\omega) \rangle$ is the final state in $\alpha(\omega)$, 
and $M=(H,L)$ and $m=(e ,h)$. 
This represents the bond correlation in the NN $ij$ bond where the $i$ and $j$ sites are identified as the configurations of $M$ and $m$. 
In the figure, $B^{mM}(\omega)$ at $\omega=0$ represents the results in the initial state of $\alpha(\omega)$, i.e. $|GS_1 \rangle$. 
It is seen that $B^{h H}(\omega=0) \sim B^{h H} (\omega \sim 3t_A) \ne  B^{h H}(\omega \sim 2t_A)$ and 
 $B^{e L}(\omega=0) \sim B^{e L} (\omega \sim 2t_A) \ne  B^{e L}(\omega \sim 3t_A)$. 
Thus, the final state in the peak B(D) is concerned in the excitation related to the hole and HS (electron and LS). 
From these results, as well as the size dependence of the peak energy, 
we identify that the peak D corresponds to the Drude component for the photo-excited electron in the thermodynamic limit. 
On the other hand, the peak B is the excitation in the hole-HS bound state. 
Because of the resonance between $|\psi_{i h\uparrow} \psi_{j H+1} \rangle$ and $|\psi_{i H+1} \psi_{j h \uparrow} \rangle$ by $t_A$, the bonding and antibonding orbitals are formed in the hole-HS bound state [Fig.~\ref{fig:fig3}(d)]. 
The peak B is owing to the excitation from the bonding to antibonding states.  
In an isolated hole-HS bound state, the excitation energy is $2t_A$ which almost agrees to the energy of the peak B. 

\begin{figure}[t]
\includegraphics[width=\columnwidth,clip]{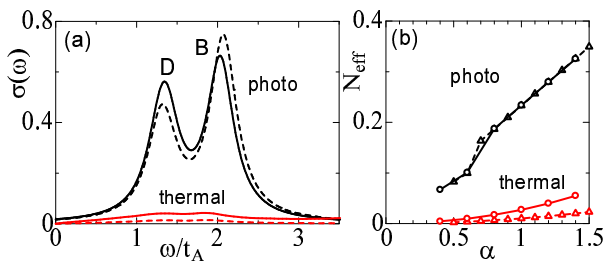}
\caption{
(a) the optical conductivity spectral at finite $T$ calculated in ${\cal H}$. 
A one-dimensional $N=5$ cluster is adopted. 
Parameters are $\Delta=8t_A$, $J=4.325t_A$ and $T=8t_A$ for the bold line 
and  $\Delta=5t_A$, $J=2.5t_A$ and $T=5t_A$ for the broken line. 
The optical conductivity spectra for the photo-excited HS state at $T=0$ are also plotted. 
(b) the effective optical carrier numbers $N_{eff}$ as functions of the transfer integral $\alpha$. 
Parameters are the same with those in (a). 
}
\label{fig:fig4}
\end{figure}
Finally, we discuss implications of the present theory for the recent pump-probe experiments in perovskite cobaltites. 
Okimoto and coworkers performed the femtosecond pump-probe spectroscopy in a series of cobaltites $R$BaCo$_2$O$_{6-\delta}$ ($R$=Sm, Gd, and Tb). These materials show the HS-metal to the LS-insulator crossover with decreasing $T$~\cite{Okimoto2}. 
The experiments in the low-$T$ LS insulating state show that 
1) 
the spectral weights appear inside of the insulating gap by photon pumping, 
but the spectral shapes are different from those in the high-$T$ HS metal, and 2) the transferred spectral weight by pumping, quantified as the effective number of carriers, increases with increasing the electron transfer integral, corresponding to increasing of the ionic radius in $R$. 
Since the photo-induced hole-HS bound state predicted in the present theory is a local object, we expect that the calculated results in small-size clusters are able to be compared qualitatively to the experimental results in the cobaltites. 
In the exact-diagonalization calculation at finite $T$, a number of HS sites $n_{HS}$ 
abruptly increases around a certain temperature $T^\ast$ which corresponds to the crossover temperature from the low-$T$ LS state to the high-$T$ HS state. 
Since the numerical value of $T^\ast(\sim t_A)$ is much higher than the observed value in cobaltites owing to the small size cluster, we do not touch $T^\ast$ itself, but focus on qualitative feature of the optical spectra above $T^\ast$. 
The optical conductivity spectra in the thermal-excited HS state as well as those in the photo-excited state are shown in Fig.~\ref{fig:fig4}(a) for the representative two parameter sets~\cite{thermal}. 
We interpret that these correspond to the experimentally observed thermally-induced and photo-induced spectral weights inside of the gap. 
We chose the temperatures where $n_{HS}$ is about 0.8 which is comparable to a value in high $T$-HS phase in LaCoO$_3$~\cite{Saitoh}. 
It is shown that the spectra in thermal-excited HS states are smaller than those in photo-excited HS state and do not show clear structure of the peaks B and D. 

Then we discuss the $R$-ion dependence of the optical spectra~\cite{Okimoto2}.  
In perovskite crystal, the ionic radius of $R$ controls the electron-transfer integral.
To simulate this effect, we introduce the parameter $\alpha$ in the transfer term in Eq.~(\ref{eq:h0}) as $t_\gamma \rightarrow \alpha t_\gamma$. 
As shown in Fig.~\ref{fig:fig1}, with increasing the transfer integral,  
the system is transferred from the phase where the LS ground state remains after photo-irradiation into the phase where HS is induced by photon. 
In Fig.~\ref{fig:fig4}(b), we present the effective number of optical carriers defined by $N_{eff}=\int^{\omega_c}_0 \sigma(\omega) d \omega$ in the photo-excited and thermal-excited HS states. 
The cut-off energy $\omega_c$ is taken to be the energy where the peak B is fully damped. 
In both the photo-excited and thermal-excited states, $N_{eff}$ increase with increasing $\alpha$. 
A number of the photo-induced effective carrier is much larger than that of the thermal-excited one.  
Through the above comparisons with the theory and the experiments, we claim that the two characteristic features suggested experimentally, i.e. 1) difference between the spectra in thermal- and photo-excited HS states, and 2) $R$ dependence of $N_{eff}$, are explained by the present calculations. 

In summary, we have studied the photo-induced spin state change in itinerant correlated electron system. 
The effective models before and after photon-pumping are derived from the two-orbital Hubbard model and are analyzed by the exact diagonalization method. 
When a photon is introduced in the LS band insulator, we found a spin-polarized bound state of photo-excited hole and HS. 
We show that this bound state directly reflects the optical pump-probe spectra. 
There is qualitative difference between the optical spectra in the photo-excited HS state and that in thermal-excited HS one. 
These results as well as the electron transfer dependence of the optical spectra well explain the recent femtosecond spectroscopy experiments in perovskite cobaltites. 

\par
Authors would like to thank 
J. Ohara, Y. Inoue, Y. Okimoto, S. Koshihara, S. Iwai and T. Arima for their valuable discussions. 
This work was supported by KAKENHI from MEXT, 
Tohoku University "Evolution" program, 
and Grand Challenges in Next-Generation Integrated Nanoscience.
YK is supported by the global COE program "Weaving Science Web beyond Particle-Matter Hierarchy" of MEXT, Japan.

\vfill
\eject

\begin{references}
\bibitem{exchange}
The inter-site interactions in Eq.~(\ref{eq:h0}) are explicitly given by 
$J_{HH}=(t_A^2+t_B^2)/(U+J)$, 
$J_{LL}=-\{4(t_A^2+t_B^2)f_A^2 f_B^2  \}/(U'-2J+2{\widetilde \Delta})$, 
$J_{HL}=-(t_A^2+t_B^2)\{f_B^2/(U'-\Delta+{\widetilde \Delta})+f_A^2/(U'+\Delta+{\widetilde \Delta}) \}$, 
$J_{++}=-2t_A t_B f_A f_B \{ 1/(U+J)+1/(U'-2J+2{\widetilde \Delta})  \}  $, and 
$J_{-+}=-2J_{HL}t_A t_B/(t_A^2+t_B^2)$ with ${\widetilde \Delta}=\sqrt{\Delta^2+J^2}$.
\bibitem{thermal}
For the photo-excited HS state, we present the absorption part of the optical conductivity spectra defined by $\alpha(\omega)/\omega$ where the optical excitations are restricted in the Hilbert space for ${\cal H}_1$. 
This is reasonable to examine the spectra in the frequency region in Fig.~\ref{fig:fig4}(a).
\end{references}
\end{document}